\documentclass[11pt,twoside]{article}
\usepackage{asp2010}

\resetcounters

\begin{document}

\title{Precision and Resolution in Stellar Spectropolarimetry}   
\author{D.M. Harrington$^1$, J.R. Kuhn$^2$} 

\affil{$^1$Institute for Astronomy, University of Hawai'i, Honolulu, HI 96822} 
\affil{$^2$Institute for Astronomy Maui, University of Hawai'i, Pukalani, HI 96768}

\begin{abstract} 

	Stellar spectropolarimetry is a relatively new remote sensing tool for exploring stellar atmospheres and circumstellar environments. We present the results of our HiVIS survey and a multi-wavelength ESPaDOnS follow-up campaign showing detectable linear polarization signatures in many lines for most obscured stars. This survey shows polarization at and below 0.1\% across many lines are common in stars with often much larger H$_\alpha$ signatures. These smaller signatures are near the limit of typical systematic errors in most night-time spectropolarimeters. In an effort to increase our precision and efficiency for detecting small signals we designed and implemented the new HiVIS bi-directionally clocked detector synchronized with the new liquid-crystal polarimeter package. We can now record multiple independent polarized spectra in a single exposure on identical pixels and have demonstrated 10$^{-4}$ relative polarimetric precision. The new detector allows for the movement of charge on the device to be synchronized with phase changes in the liquid-crystal variable retarders at rates of $>$5Hz. It also allows for more efficient observing on bright targets by effectively increasing the pixel well depth. With the new detector, low and high resolution modes and polarization calibrations for the instrument and telescope, we substantially reduce limitations to the precision and accuracy of this new spectropolarimetric tool.  \\

\end{abstract}

\section{Introduction}

	Spectropolarimetry has been used in many fields as a diagnostic tool for deriving properties of astrophysical fluids. There are many types of spectropolarimetric models and observed polarized line profile morphologies across the HR diagram. The mechanisms invoked to explain these profiles include scattering, emission, absorption, atomic population imbalances (pumping) and potential modification of all these by magnetic fields.
	
	In the context of circumstellar material, there are a few scattering models that are commonly invoked. The depolarization morphology where unpolarized line emission dilutes a polarized stellar continuum was developed in the Be star context and requires an asymmetric circumstellar envelope for creating continuum polarization via electron scattering and dilution of the polarized continuum by unpolarized line emission. There are line polarization effects that can arise from doppler shifts induced when scattering off thin circumstellar disks. The asymmetric circumstellar envelope and electron-scattering concept has been used to model or constrain Be circumstellar environments by fitting spectropolarimetric observations and also comparing polarization measurements to interferrometric measurements. There are also codes to produce synthetic polarized line profiles and continuum polarization for a hot star winds. 
		
	There have been a number of linear spectropolarimetric studies performed to date at spectral resolutions of a few to several thousand. Depolarization effects have been used in interpreting spectropolarimetric line profiles in many stars including Herbig Ae/Be, Be, B[e] and O-supergiants. The disk-scattering model was used to discuss various types of line profiles in TTauri and Herbig Ae/Be stars, O-type stars, as well as hot-massive stars. Wolf-Rayet stars have been observed and interpreted using these scattering frameworks as have massive post-red supergiants.

\begin{figure*}
\begin{center}
\includegraphics[width=0.35\linewidth, angle=90]{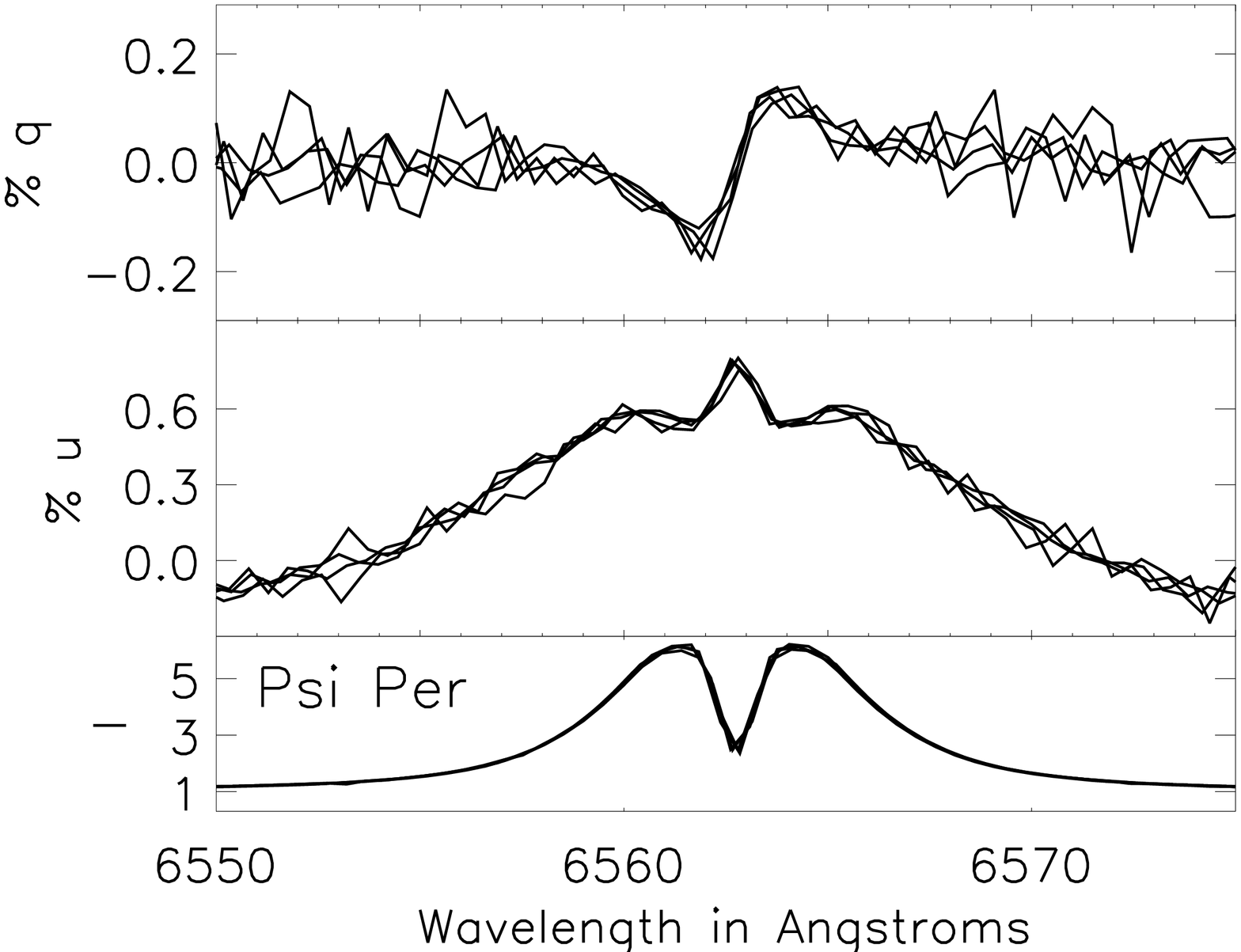}
\includegraphics[width=0.35\linewidth, angle=90]{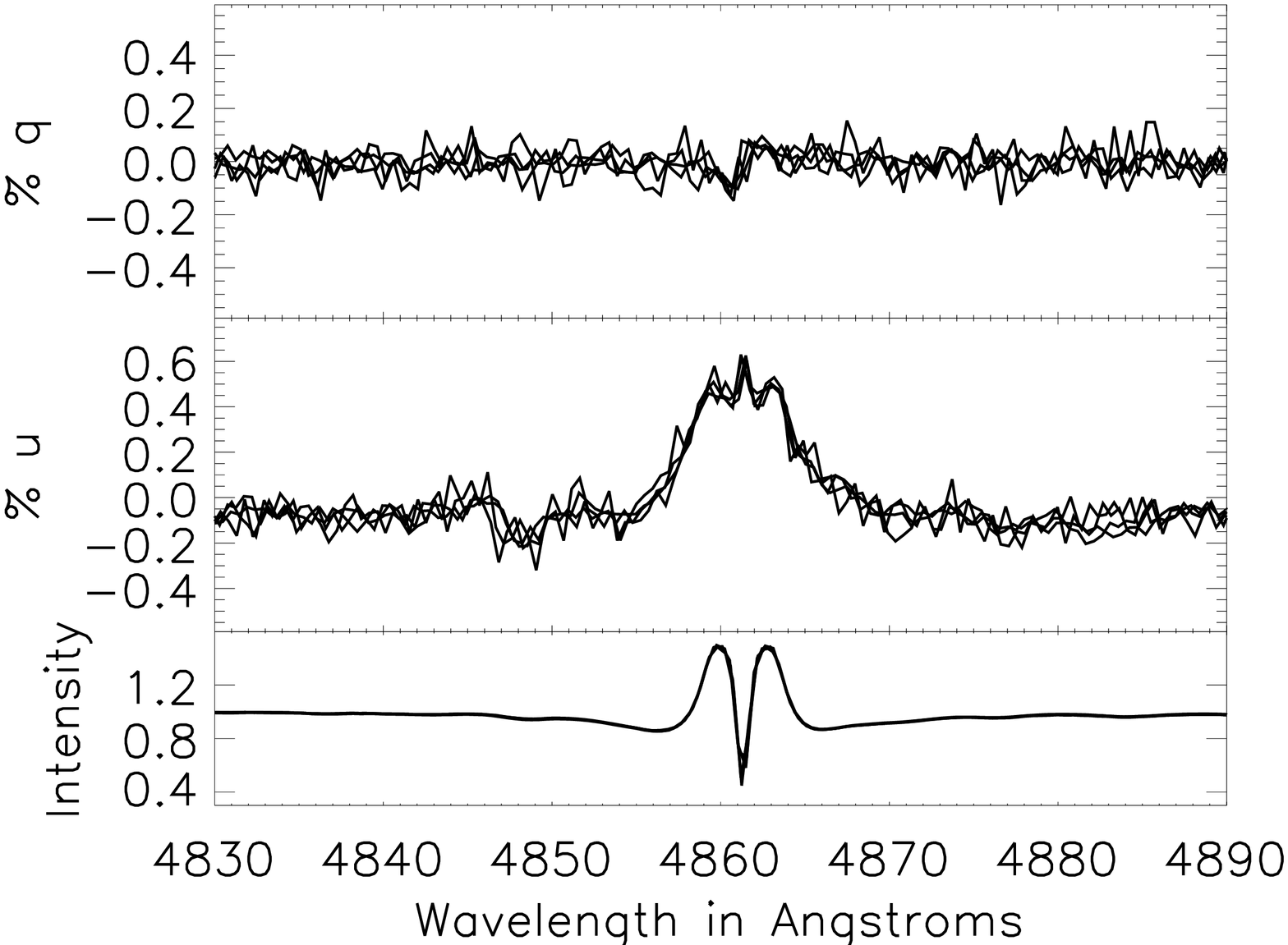} \\
\includegraphics[width=0.35\linewidth, angle=90]{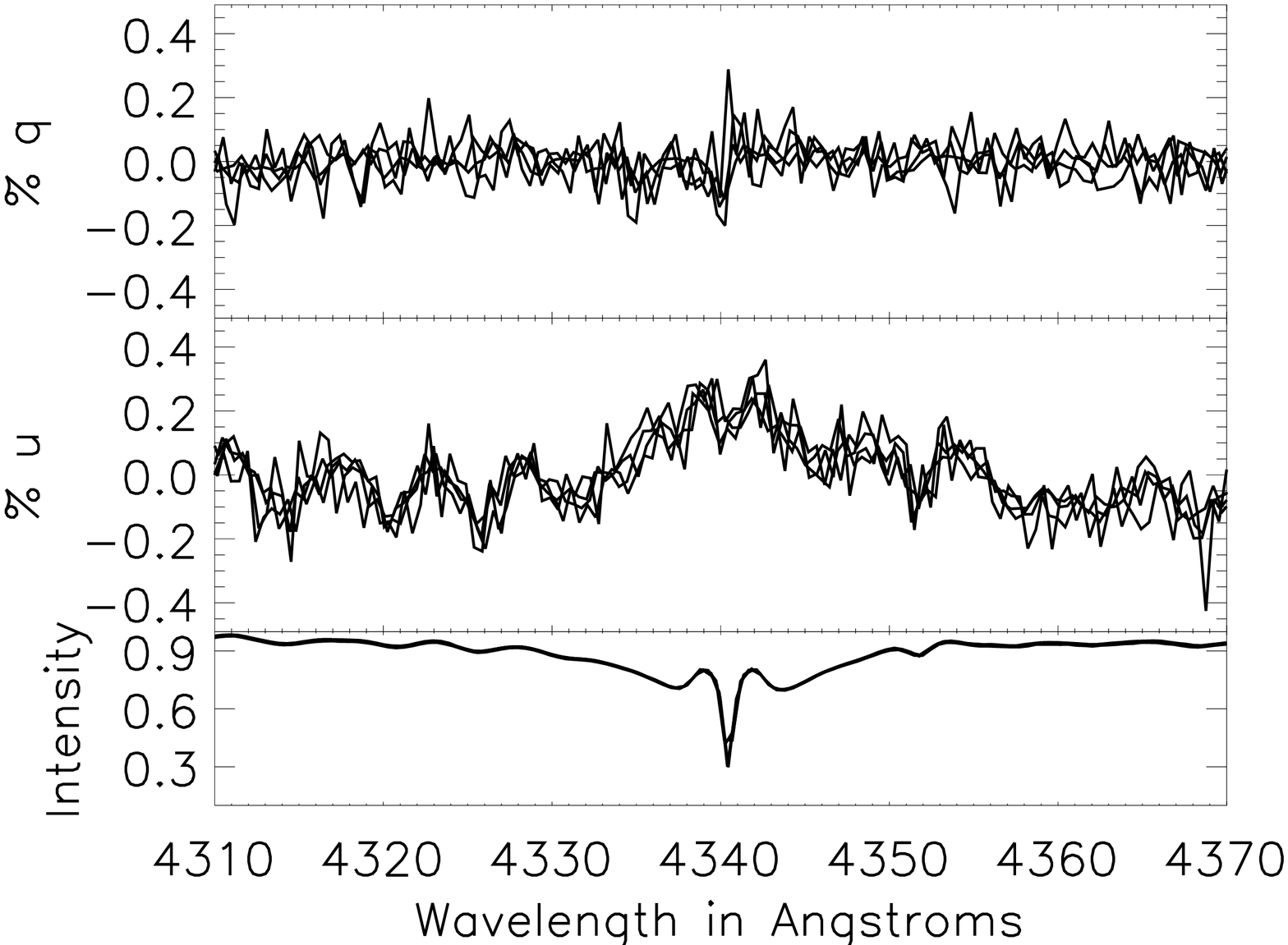}  
\includegraphics[width=0.35\linewidth, angle=90]{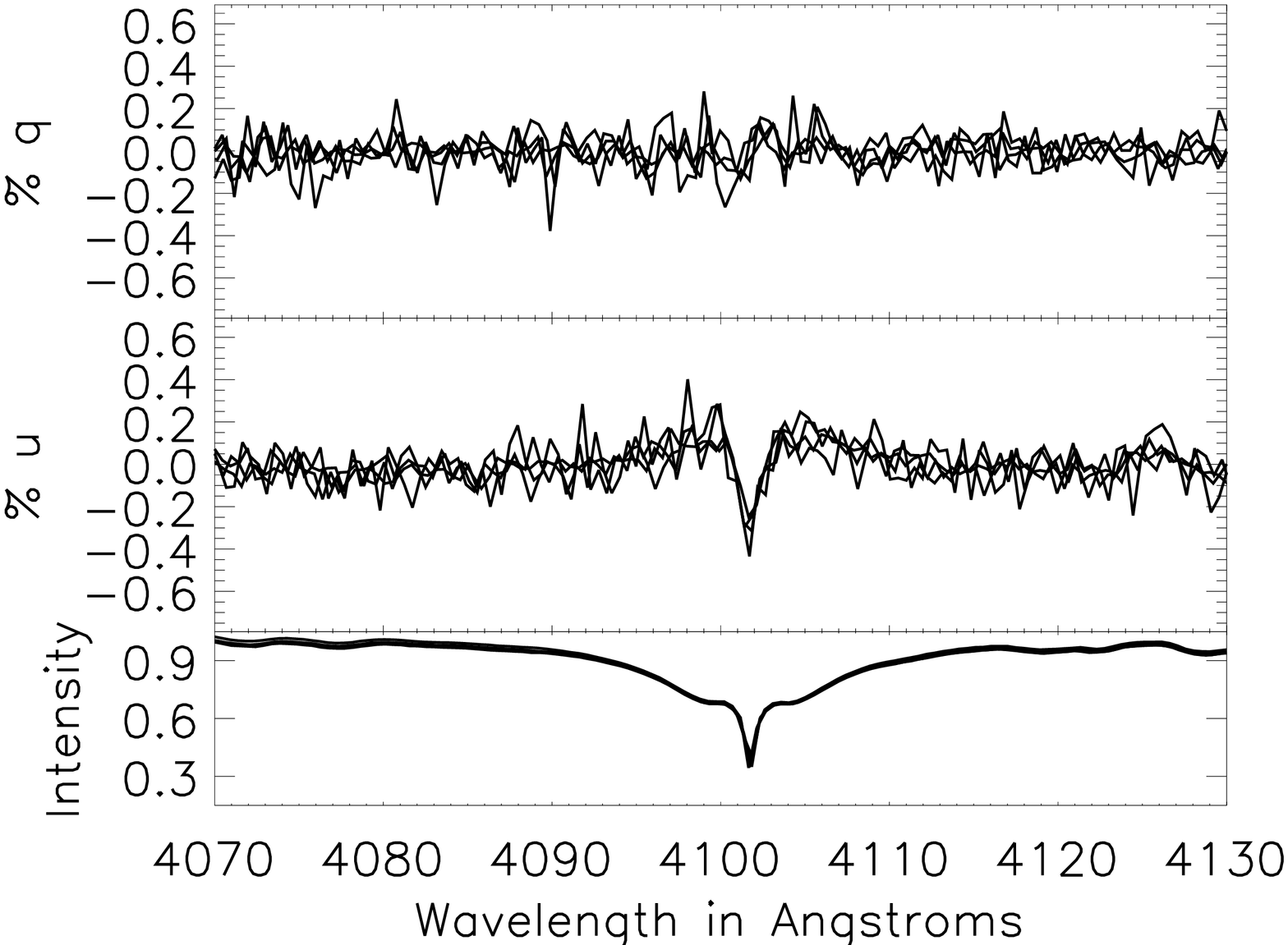} 
\caption{The ESPaDOnS spectropolarimetry for $\psi$ Per in the H$_\alpha$ H$_\beta$ H$_\gamma$, H$_\delta$ lines for multiple epochs. \label{hbalmer} }
\end{center}
\end{figure*}

	A review of the large number of papers outlining the scattering models and observations is beyond the scope of this proceedings, but some general features can be mentioned. Most of the observations presented in the observations were taken at low to moderate spectral resolution with various resolutions and binning procedures, with typically less than 10 independent polarization measurements across a spectral line. In addition, most studies use a few hours to a few nights worth of observations. In a series of papers, we outlined an H$_\alpha$ survey of many stars at high spectral resolution we have performed on over 100 observing nights (\citealt{har07, har08, har09a, har09b}). These observations found several examples of more complex line profile morphologies. Many of the Herbig Ae/Be observations showed polarization line profiles where only the absorptive component of a P-Cygni profile was polarized differently than the adjacent stellar continuum. The emission line polarization matched the continuum, ruling out the depolarization or typical scattering mechanisms. 

	\citealt{kuh07} introduced optical pumping (atomic alignment) as a mechanism for interpreting H$_\alpha$ spectropolarimetric line profiles from circumstellar material. This and related mechanisms are known in the solar community as atomic alignment (cf. \citealt{tru97}, \citealt{tru02}, \citealt{cas02}, \citealt{tru07}, \citealt{bel09}). It has also been called Zero-field Dichroism (\citealt{man03}). \citealt{ase05} apply this concept to circumstellar masers. Atomic alignment effects have been used in interpreting solar H$_\alpha$ observations (cf. \citealt{lop05}). A related computational method for a similar effect was explored by \citealt{yan06, yan07, yan08}. There the alignment of atoms by magnetic-field-modified emission is explored while the optical pumping effect of \citealt{kuh07} is independent of magnetic fields.

\section{The HiVIS Survey \& ESPaDOnS follow-up}

	 The High-resolution Visible and Infrared Spectrograph (HiVIS) is a spectropolarimeter on the 3.67m Advanced Electro-Optical System (AEOS) telescope with resolutions between 10,000 and 50,000. HiVIS has both achromatic wave plates and liquid crystal variable retarder (LCVR) full-Stokes modes and a dedicated IDL data reduction package (Harrington et al. 2006, Harrington \& Kuhn 2008, 2010). We have performed a $>$100 night survey with HiVIS and have many detections of linear polarization signals associated with H$_\alpha$ line profile. With high spectral resolution (R$>$10000) there are comparatively large amplitude (0.2-2\%) signals in Herbig Ae/Be, Be, PostAGB and other emission-line stars presented in \citealt{kuh07}, \citealt{har07, har08, har09a, har09b}. The absorptive polarization effects are ubiquitous in the H$_\alpha$ line of Herbig Ae/Be stars with roughly 2/3 of the stars showing this linear polarization signature. The classical Be stars more typically showed a broad `depolarization' morphology (10/30). About half of these show additional antisymmetric signatures in the absorptive part of the line profile (4/10) and another sub-set (5/30) show complex linear polarization spectral morphologies. The presence of absorptive polarization effects in most of the obscured stars in this survey suggests that the phenomenon may be present in any obscured star. 
	 
	The follow-up observations for this project were performed with the 3.6m Canada France Hawaii Telescope (CFHT) using the ESPaDOnS spectropolarimeter with a nominal average spectral resolution of $R=\frac{\lambda}{\delta\lambda}=68000$ (cf. \citealt{don97, don99}). The Canadian Astronomy Data Centre (CADC) also provides archival access to all ESPaDOnS data which was searched for additional observations. All observations were reduced with the dedicated script, Libre-ESPRIT (\citealt{don97}). 
	
\begin{figure*}
\begin{center}
\includegraphics[width=0.24\linewidth, angle=90]{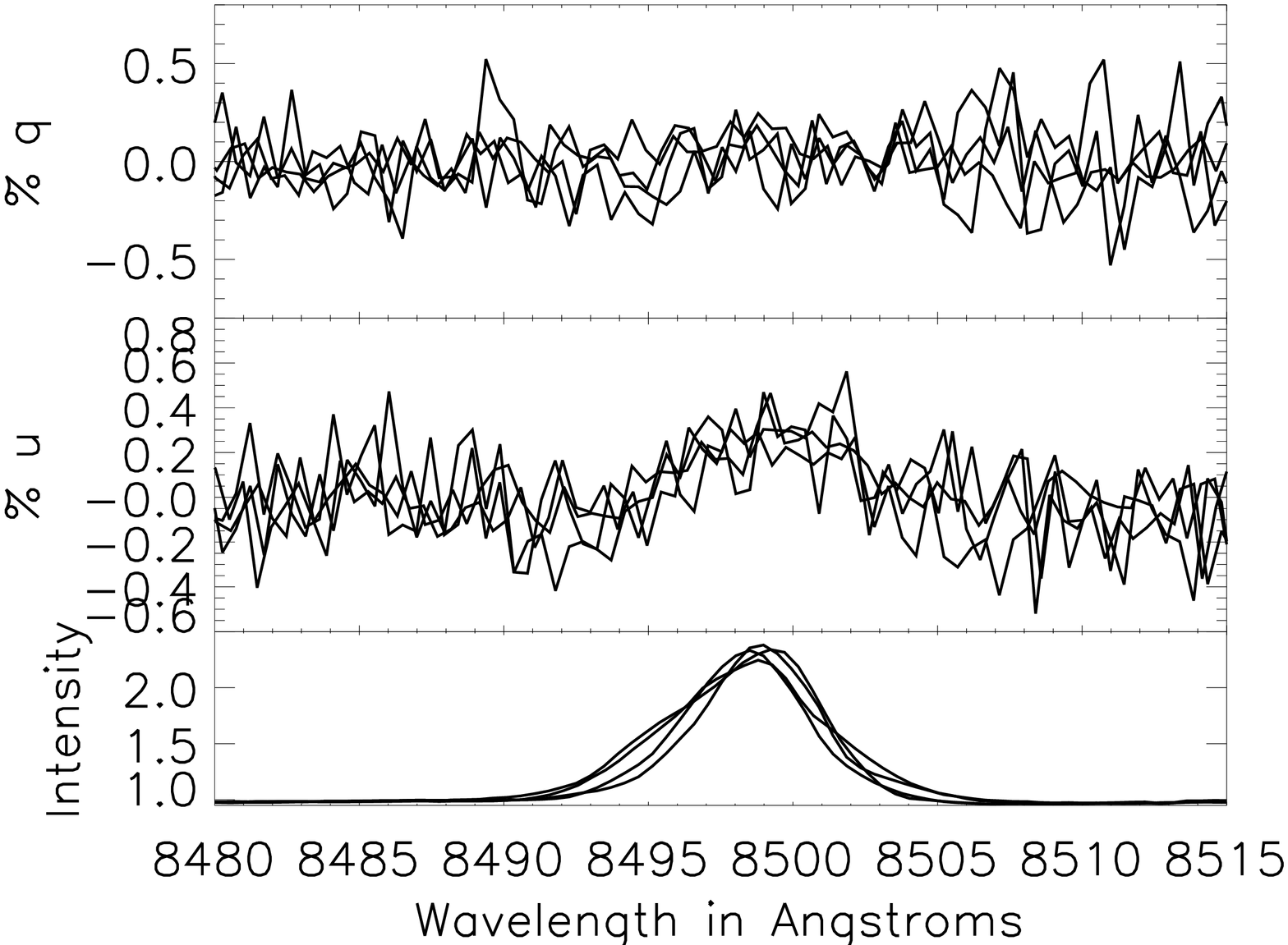} 
\includegraphics[width=0.24\linewidth, angle=90]{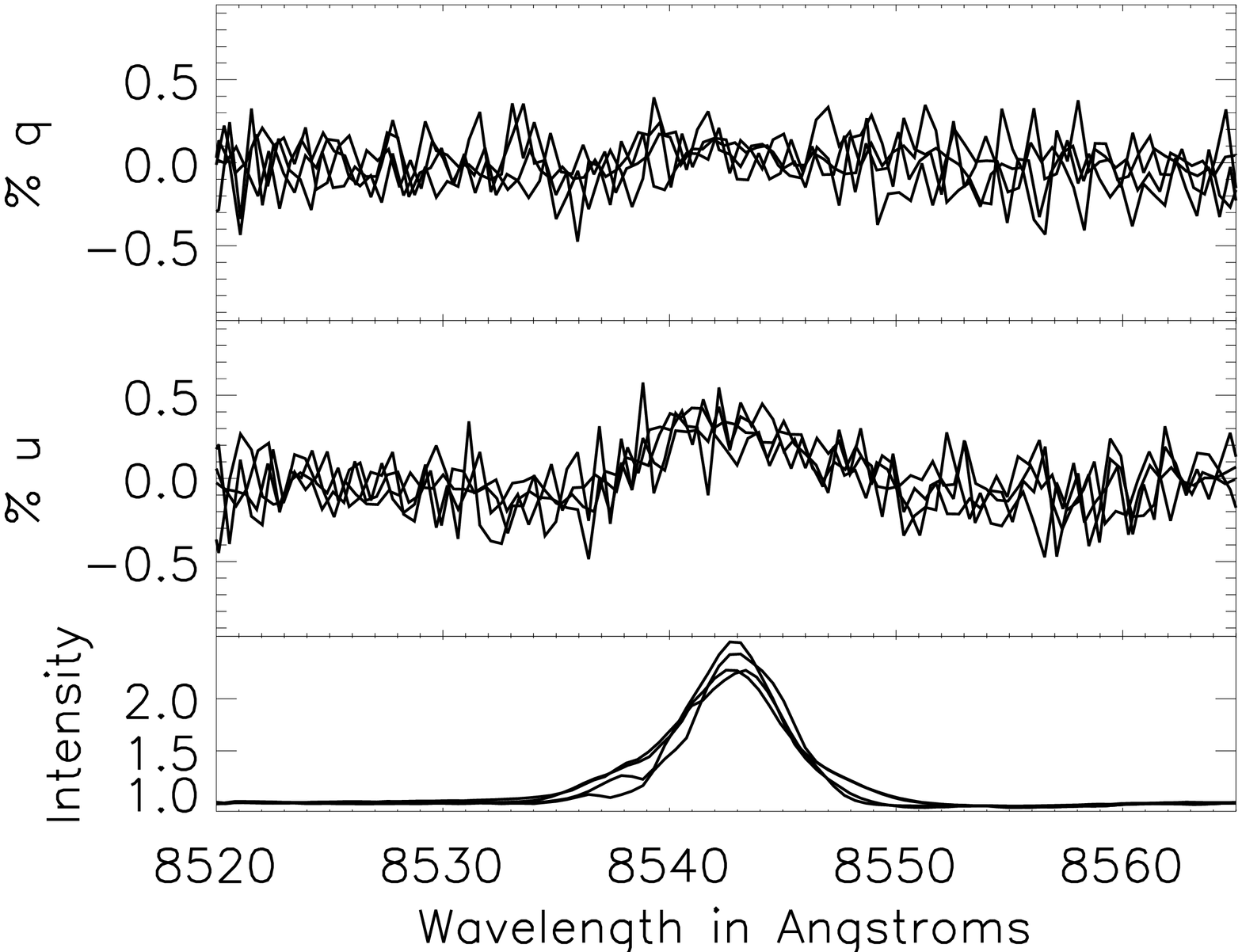}
\includegraphics[width=0.24\linewidth, angle=90]{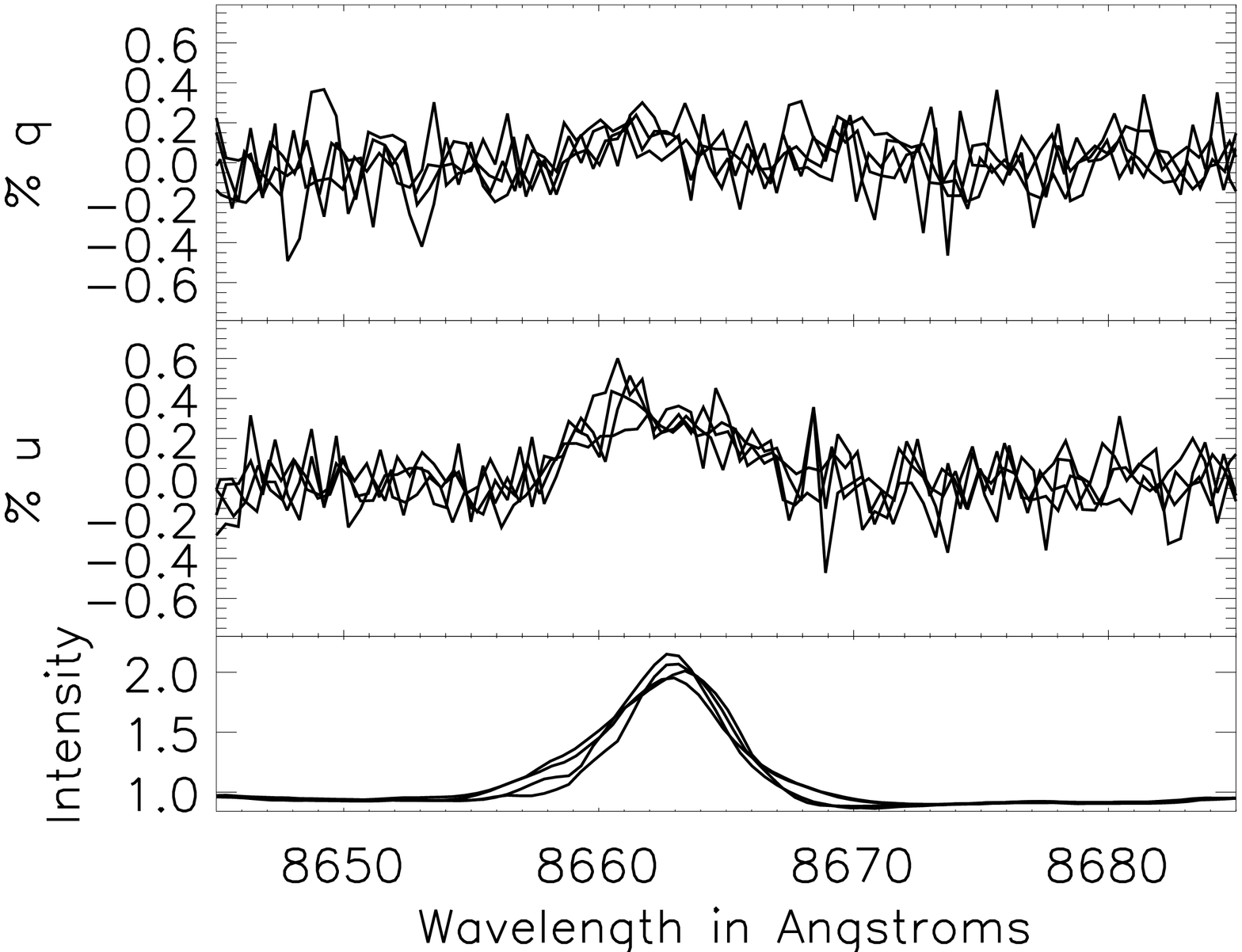} \\
\includegraphics[width=0.24\linewidth, angle=90]{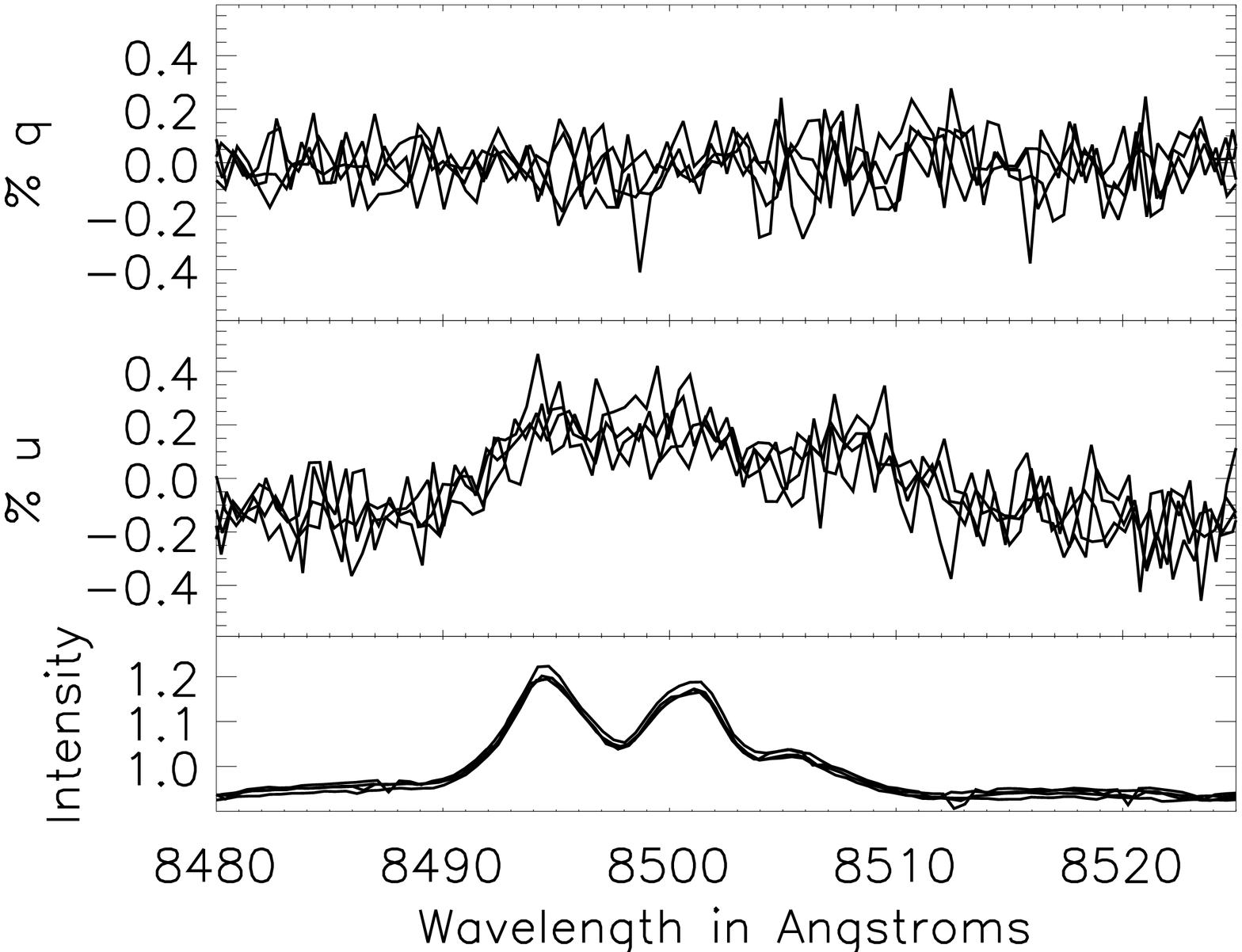} 
\includegraphics[width=0.24\linewidth, angle=90]{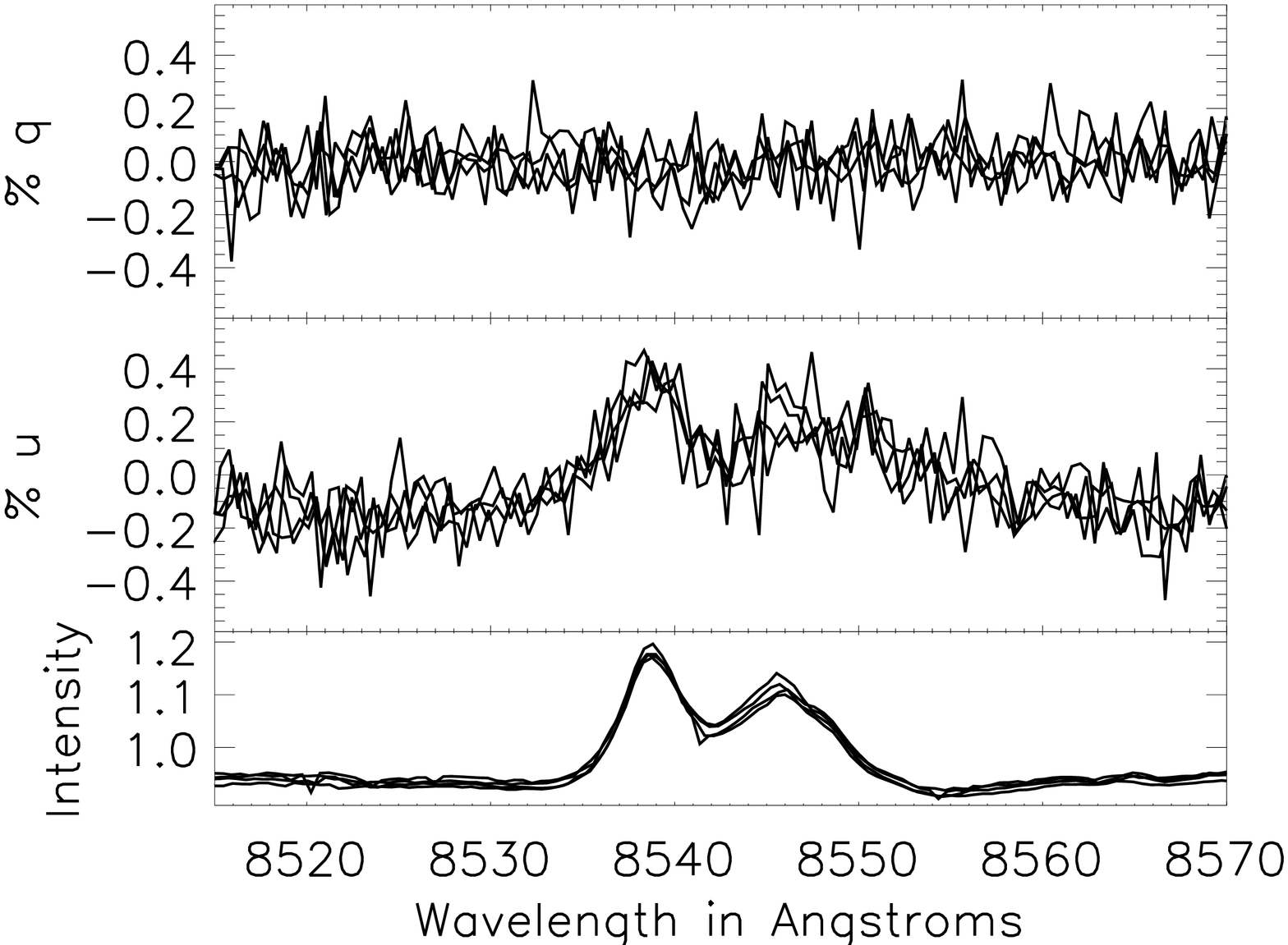}
\includegraphics[width=0.24\linewidth, angle=90]{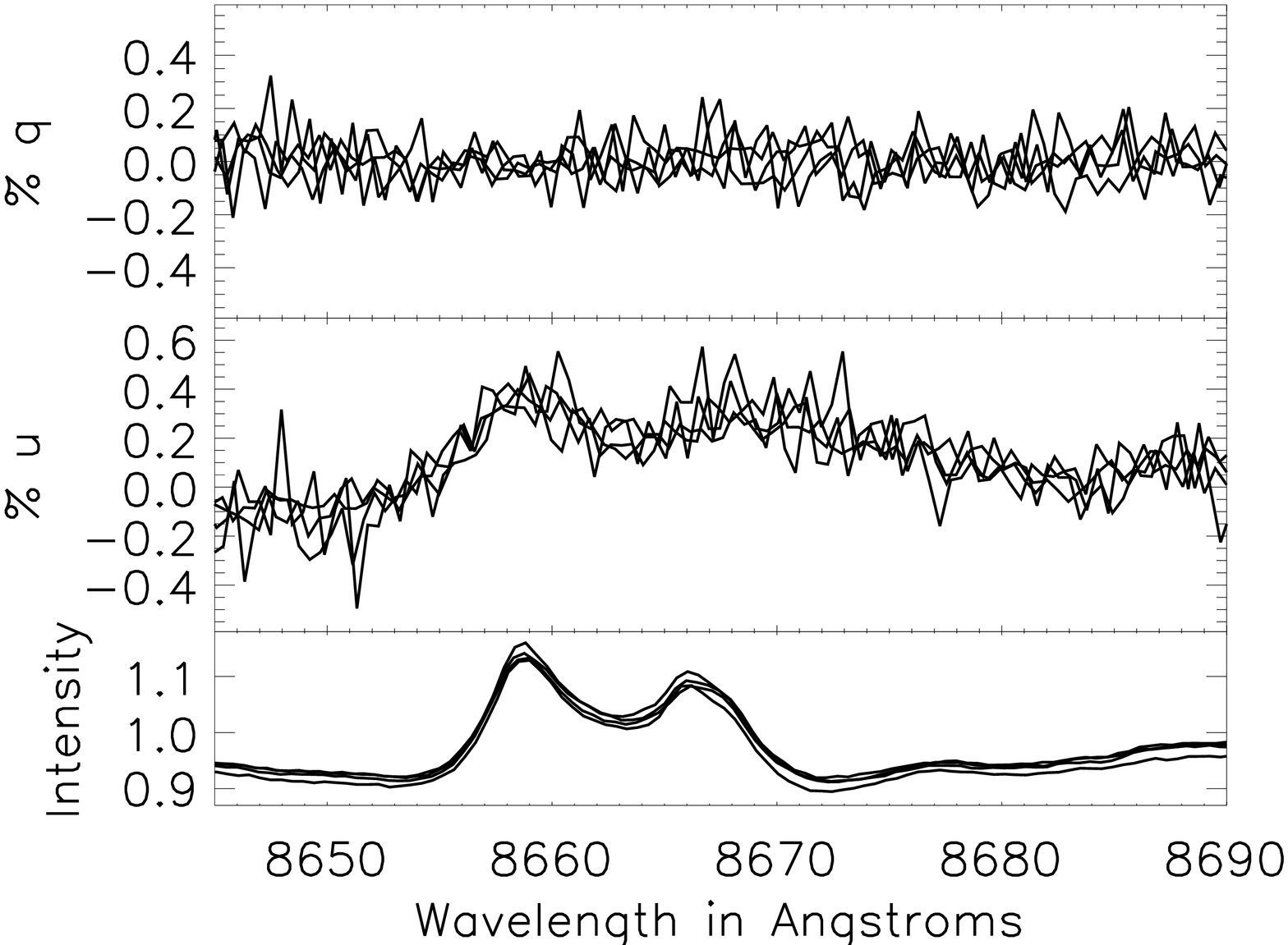} \\
\caption{The spectropolarimetry for MWC 480 and Psi Per in the Ca near-infrared triplet lines 850, 854 and 866nm. \label{caNIR}}
\end{center}
\end{figure*}

	The ESPaDOnS targets were known detections from \citealt{har09a} and \citealt{har09b} as well as supplementary archive observations. Our present sample includes 18 HAeBe stars, 8 Be stars, 9 Post-AGB stars and 8 other emission-line stars. We commonly find spectropolarimetric signatures in 15 lines at or above the 0.1\% amplitude: H$_\alpha$, H$_\beta$, H$_\gamma$, H$_\delta$, Na D lines, Ca NIR Triplet as well as Fe, He and Ox lines. As an example, Figure \ref{hbalmer} shows the H$_\alpha$ - H$_\beta$ - H$_\gamma$ - H$_\delta$ sequence for the Be star $\psi$ Per. There is substantial change in the line profile morphology and amplitude for each of the lines.  Figure \ref{caNIR} shows examples of the Ca near-infrared triplet lines at 850nm, 854nm and 866nm. The detections are lower amplitude but the detections still show some morphological variations. With high spectral resolution and high precision detections across multiple lines, much more quantitative constraints on stellar environments have become possible.

\section{New Charge-Shuffling Detector}

	For many targets it is calibration of the telescope and instrument systematic errors that limit the polarimetric precision of an observation. Moving optical elements, detector effects such as non-uniform sensitivity or nonlinearity, unstable optical beams, atmospheric seeing and transparency variations can all induce systematic errors that can be much larger than the photon noise. Efforts to stabilize the instrument and telescope, use non-moving optics, calibrate and characterize the system are essential when observing small amplitude signatures. 

\begin{figure*} [!h, !t, !b]
\begin{center}
\includegraphics[width=0.35\linewidth, angle=90]{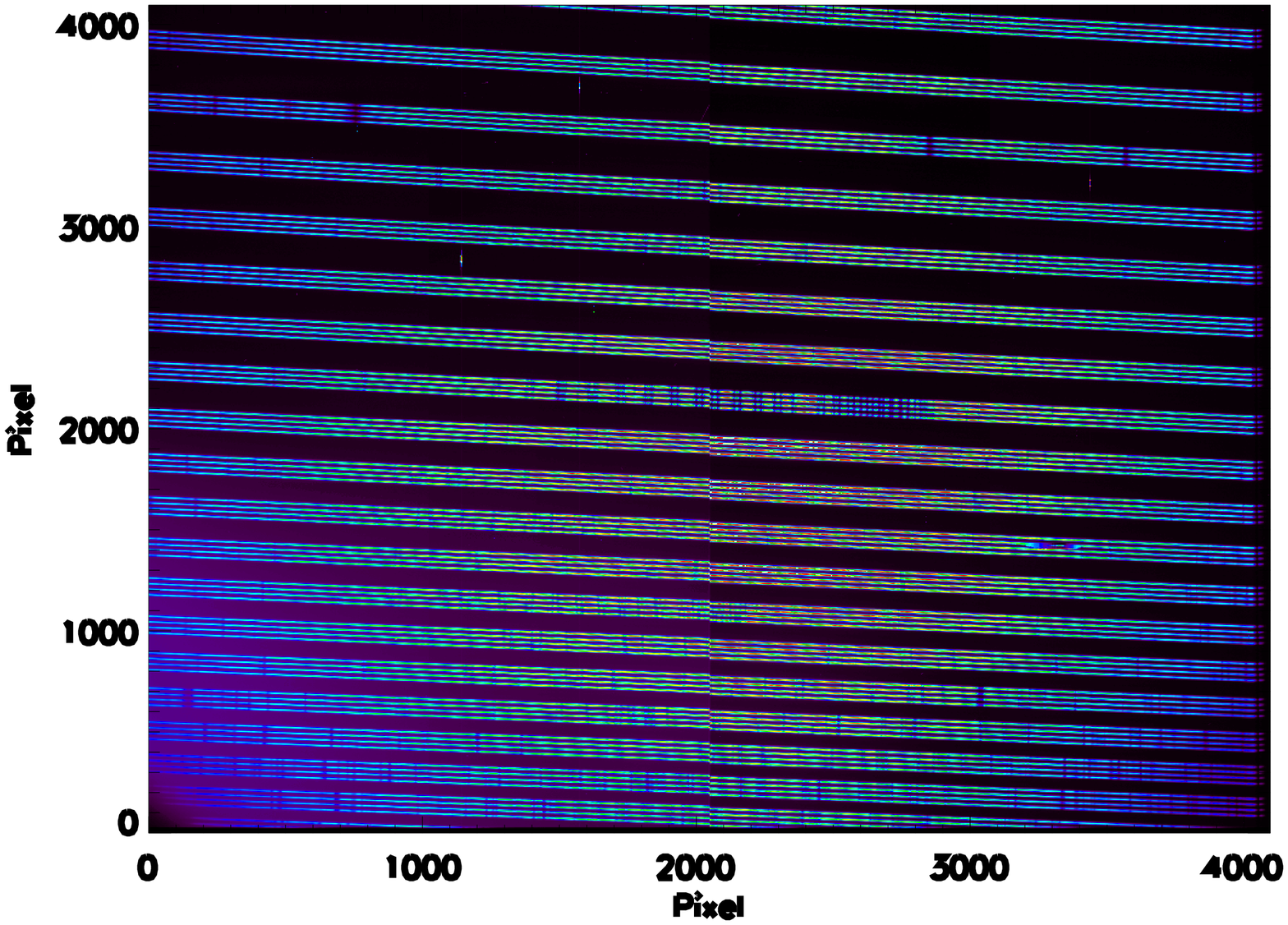}
\includegraphics[width=0.35\linewidth, angle=90]{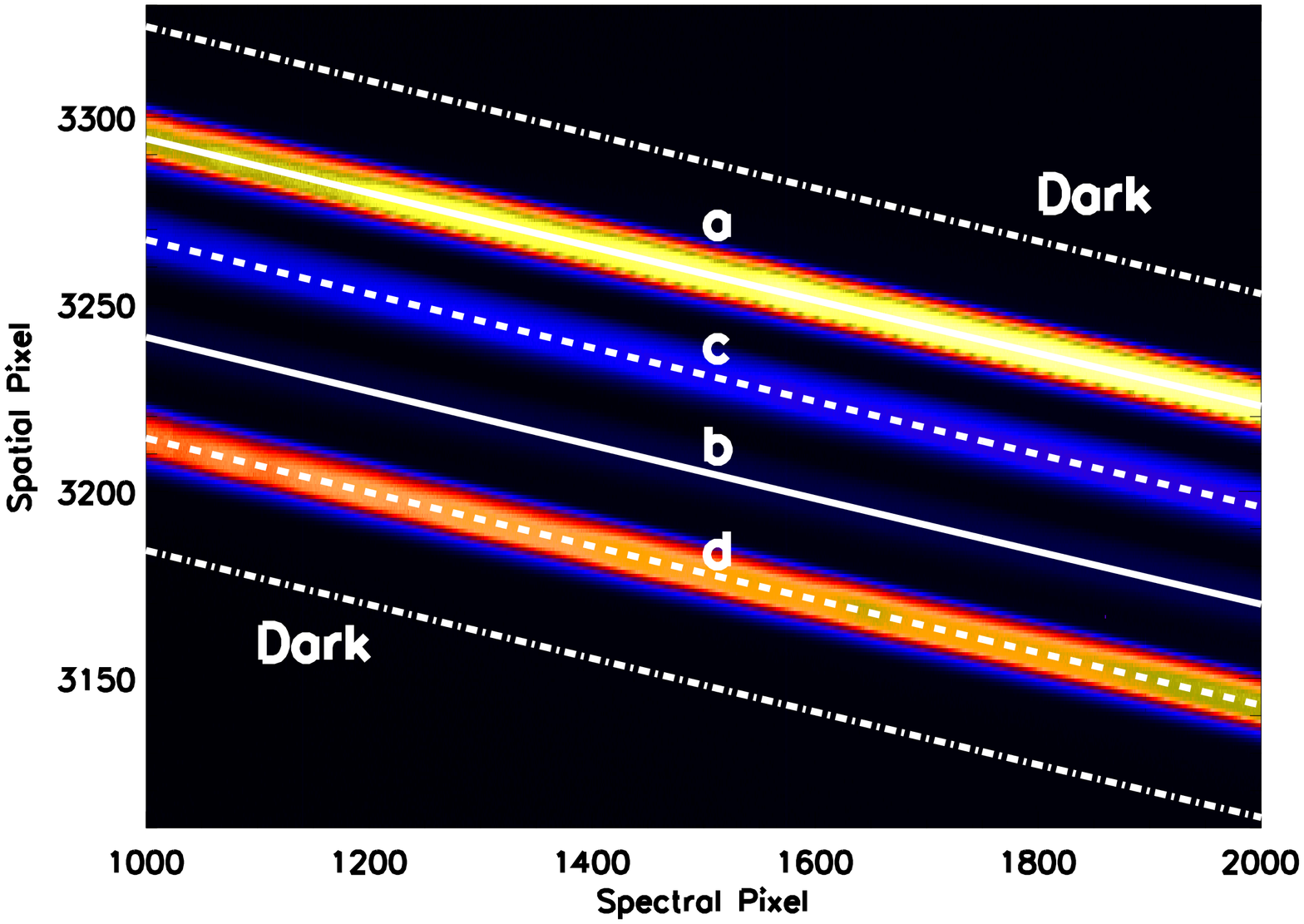}
\caption{ \label{rawcsd} The left hand panel shows a first-light Charge-Shifting Detector image in 'weave' mode of HR 3980. The right hand panel shows a calibration frame illustrating the extracted orders. The polcal stage is feeing in pure +Q. The LCVR's are set to measure $\pm$Q. The image is linearly scaled from 0 to 32,000 DN to show the bright orders as orange and the dark orders as blue/black. }
\end{center}
\end{figure*}

	 High-resolution night time astronomical spectropolarimeters have many common design elements and are almost always dual-beam systems with rotating achromatic retarders. Detector errors must be minimized by differencing signals on identical pixels given night-time observing constraints. Rotating achromatic retarders are placed before a calcite analyzer such as a Wollaston prism or a Savart plate. ESPaDOnS and Narval use two half-wave and one quarter-wave Fresnel rhombs before a Wollaston prism. Two other dual-beam instruments, PEPSI on the 8.4m LBT at R$\sim$310,000 and HARPS on the ESO 3.6m telescope at R$\sim$115,000 are in various stages of construction (\citealt{str03, str08}, \citealt{sni08, sni10}). The HARPS polarimetric package required the analyzer to be a Foster prism with separate quarter-wave and half-wave super-achromatic plates for circular and linear polarization that cannot be used simultaneously and must change between observations. The PEPSI design is similar to ESPaDOnS in that a lens collimates the beam before the retarders and a Wollaston analyzer. Instead of Fresnel rhomb retarders, a super-achromatic quarter-wave plate is chosen and linear polarization sensitivity is achieved only by physically rotating the entire polarization package. With HiVIS, a Savart plate is used as the analyzer after either rotating achromatic wave plates or LCVRs.
 
	 HiVIS is somewhat unique in that there are many oblique reflections in the optical path before the analyzer. The reflections cause quite severe cross-talk but the induced polarization and depolarization are almost always less than 5\%. The detected spectropolarimetric signatures are thus difficult to disentangle, but both detections and non-detections are significant and the spectropolarimetric morphology is preserved across an individual spectral line. We have found that effectively the cross-talk can be approximated well as a series of rotations in quv space. 
 
\begin{figure} [!h, !t, !b]
\begin{center}
\includegraphics[width=0.32\linewidth, angle=90]{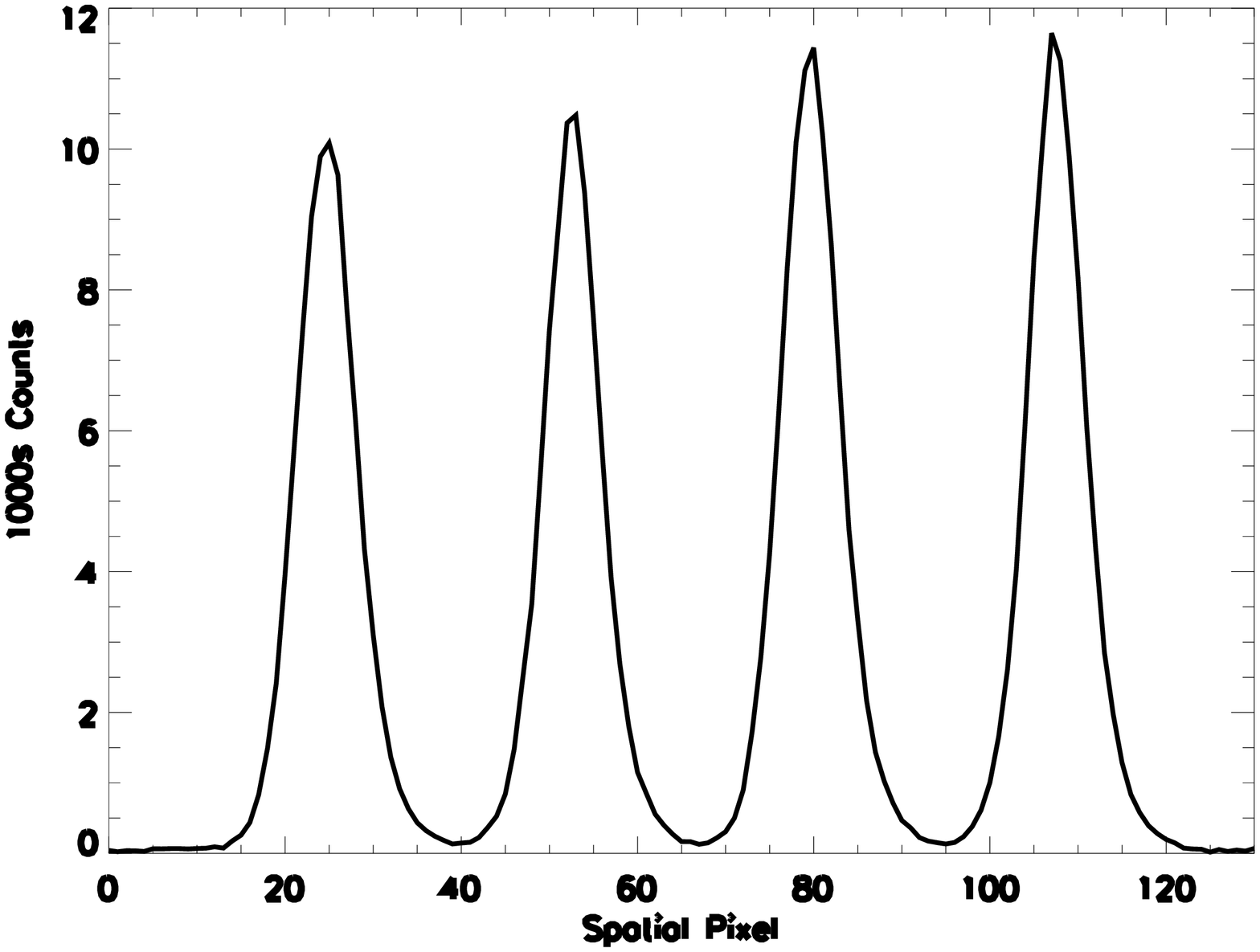}
\includegraphics[width=0.32\linewidth, angle=90]{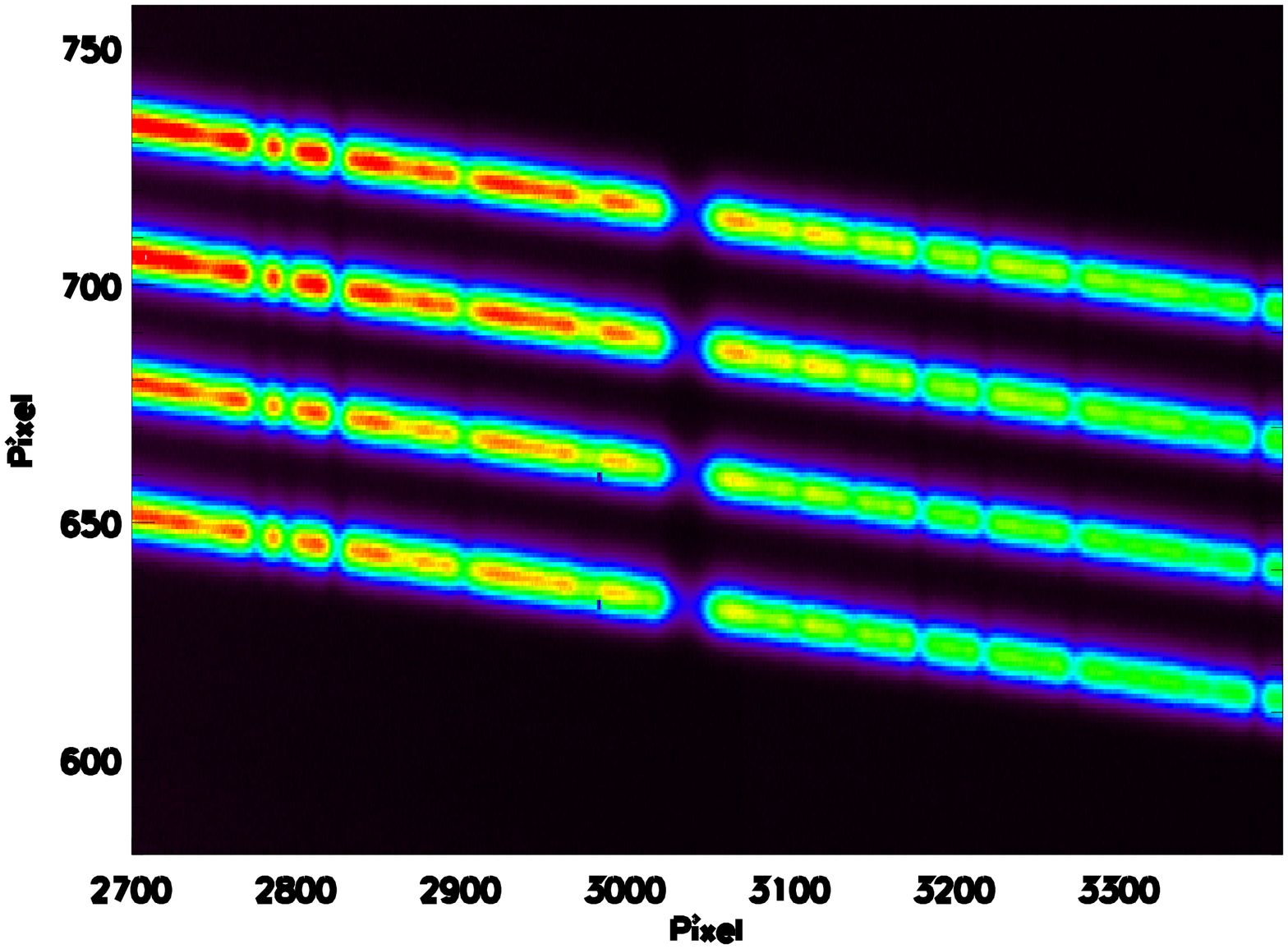} \\
\includegraphics[width=0.32\linewidth, angle=90]{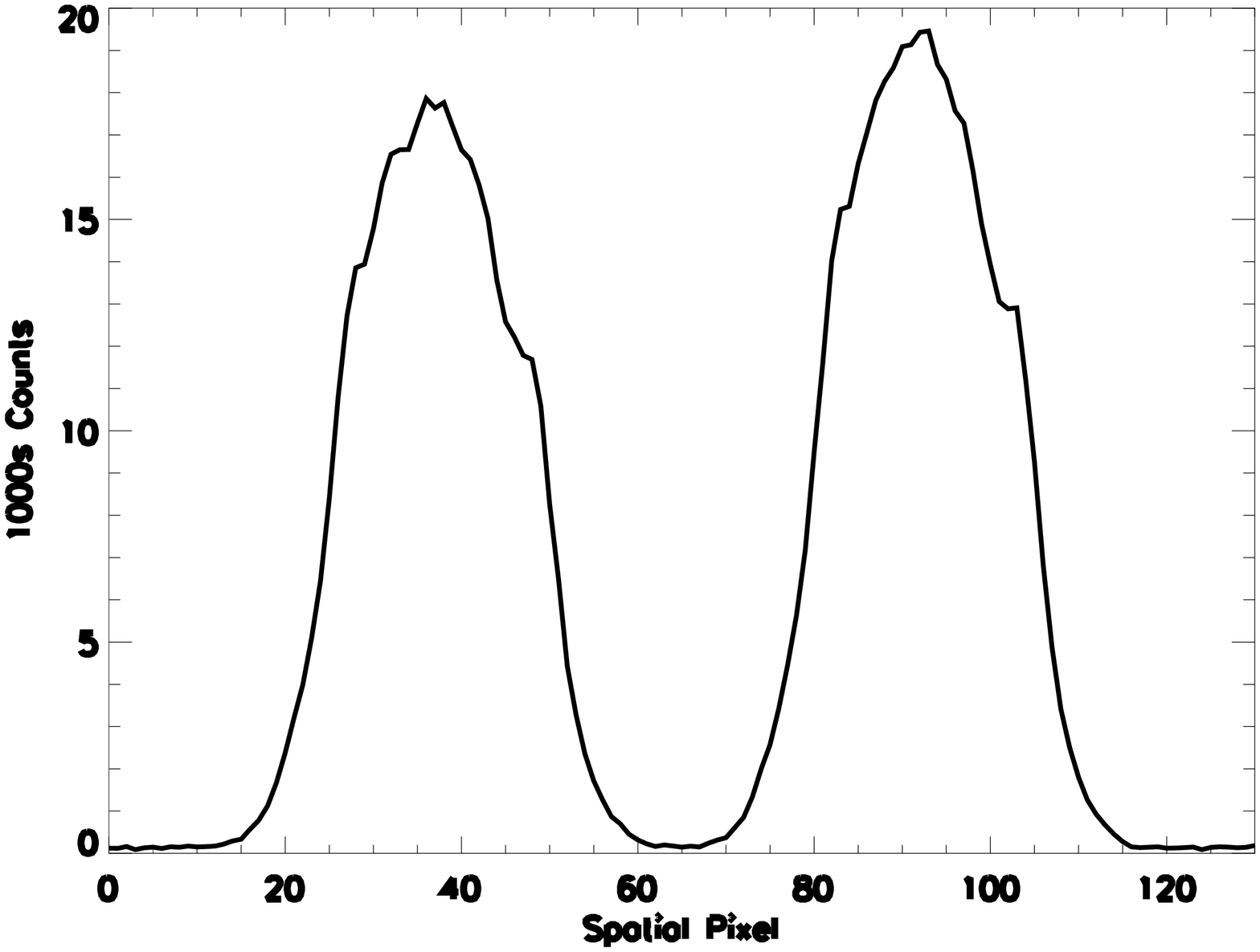}
\includegraphics[width=0.32\linewidth, angle=90]{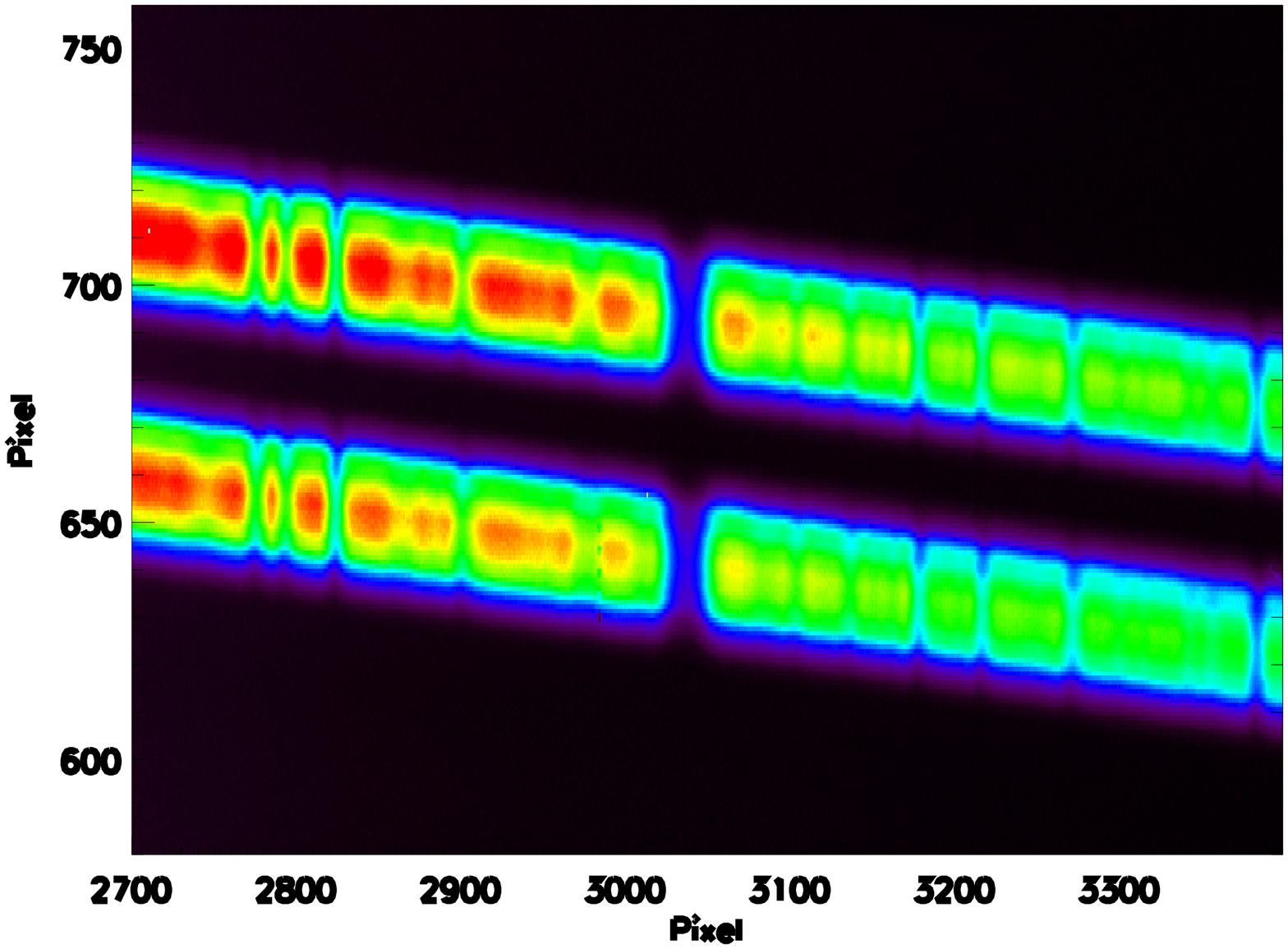} \\
\caption{ \label{csdmodes} This Figure shows the first-light sequences for the new detector in 'weave' and 'smear' modes. Left side panels show the point-spread function while right side panels show a small region of the detector. The first row shows the first-light 'weave' routine where the spectra are shuffled by half the Savart plate separation to allow spectral orders to inter-lace. The second row shows the 'smear' routine where a 6-position loop is run during integration to spread charge to adjacent pixels to increase the effective well-depth of the pixels. }
\end{center}
\end{figure}

	In an effort to remove systematic errors, many instruments use rapid modulation of the incident beam polarization state synchronous with charge motion the detector to remove time-dependent systematic effects. In some instruments, the temporal modulation is used to record multiple individual Stokes parameters at the same time. Other instruments use the temporal modulation to build up charge recording different polarization states on identical pixels. If this temporal shifting exceeds a kilohertz, then seeing errors can be minimized. For instance, the various incarnations of the ZIMPOL I, II and III solar imaging polarimeter have used piezo-elastic modulators or ferro-electric liquid crystals in combination with charge shuffling on a masked CCD to remove seeing induced systematic errors (cf. \citealt{gan04}, \citealt{pov01}, \citealt{ste07}). The Advanced Stokes Polarimeter (ASP) and La Palma Stokes Polarimeter (LPSP) are other notable polarimeters (c.f. \citealt{elm92}, \citealt{lit96}). This technique has been adapted for night-time spectropolarimetric use at the Dominion Astrophysical Observatory using ferro-electric liquid crystals and a fast-shuffling unmasked CCD (Monin, Private Communication).

	We have adapted the two CCID20 arrays of the HiVIS detector to now include bi-directional parallel clocking synchronized with LCVR phase changes. The Pan-STARRS group has developed a detector controller hardware and software package called STARGRASP that allows for user control of the charge motion (cf \citealt{ona08}, \citealt{bur07}, \citealt{ton97, ton08}). The detector is oriented so the charge motion along the parallel clock corresponds effectively to the spatial direction of the recorded spectra. The Savart plate produces sufficient beam displacement between orthogonally polarized beams that a dekked slit length of 1.5`` allows the detection of four polarized spectral orders on the CCD with clear separation between beams. The new observing mode allows us to accumulate charge from two different LCVR settings in four beams on the same two groups of pixels within a single exposure.

\begin{figure*} [!h, !t, !b]
\begin{center}
\includegraphics[width=0.32\linewidth, angle=90]{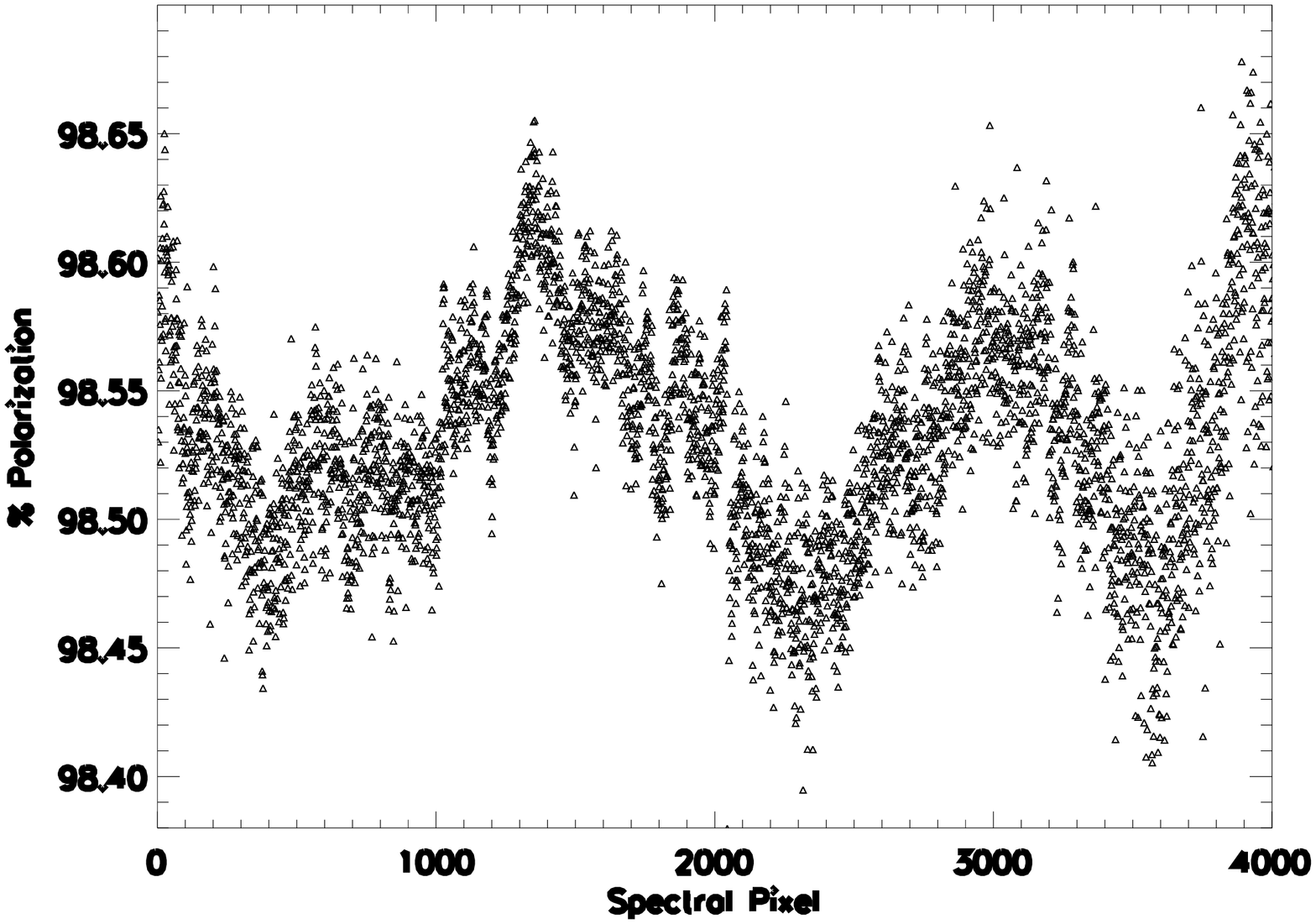}
\includegraphics[width=0.32\linewidth, angle=90]{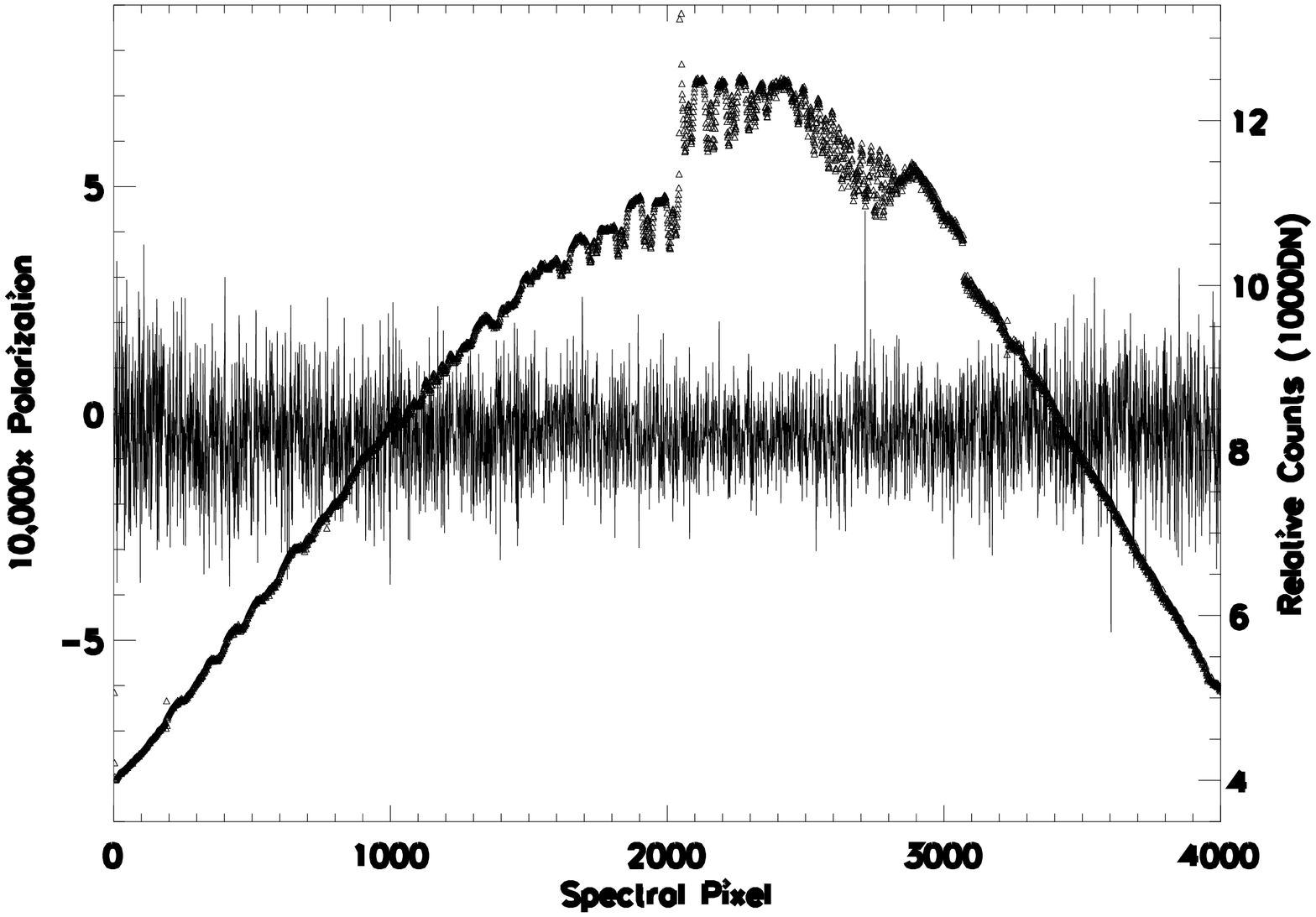}
\caption{ \label{noiseavg} The left hand panel shows the detected polarization across a single spectral order averaged for 400 individual exposures. The right hand panel shows the measured residual polarization when 200 exposures are calibrated and averaged. }
\end{center}
\end{figure*}
	
	The basic sequence is to record light for some time interval with one setting of the LCVRs, shuffle the charge to the unilluminated region of the detector, switch the LCVR voltages and to record light for another time interval.  This loop is repeated until enough charge has accumulated that a detector readout is desired. We term this the 'weave' mode since the spectral orders are interlaced. Figure \ref{rawcsd} illustrates this observing mode. The left hand panel shows a full frame recording a stellar spectrum.  The right hand panel shows a small region of the detector during a calibration exposure.  The actively illuminated pixels are represented by beams a and b. The charge accumulated from these beams is shuffled down to the region labeled c and d while the LCVRs are switched.  In this illustration, pure linear polarization was input with the HiVIS polarization calibration optics so that beams a and d have $\sim$30,000 counts while beams b and c only have $\sim$1,000 counts. 
	
	First-light with this detector was achieved in March of 2008. Figure \ref{csdmodes} shows the spatial profile (effectively the point-spread function) and a small region of the detector during a 'weave' sequence in the top row. The bottom row of this Figure shows what we call a 'smear' sequence where only two spectra are recorded but the effective well-depth of the pixels is increased by shuffling charge around during an exposure. This increases the efficiency of observing on bright targets by utilizing the detector area more efficiently and reducing the time lost to reading out the devices.

	Rotating optics as well as seeing and telescope pointing drift change the illumination pattern on the detector. This combined with imprecise calibrations can lead to systematic errors that limit the precision of any measurement. For instance, Figure \ref{noiseavg} shows on the left the polarization calibrated calculated with LCVRs at a single voltage setting. Systematic errors at the 0.1\% level are present even over small wavelength ranges. However, when the calibrations are applied to differences on identical pixels, we achieve 10$^{-4}$ precision in a single spectral pixel with no systematics seen. This is shown in the right panel of Figure \ref{noiseavg} along with the calculated intensity spectrum.

\section{Summary}

	There are many obscured stars across the HR diagram that show 0.1\% to over 1\% linear polarization signatures in many spectral lines of multiple atomic species at spectral resolutions above 10,000. We have performed a large survey of many stars in H$_\alpha$ and have followed up on these targets with ESPaDOnS to obtain detections across many spectral lines at 68,000 spectral resolution. In an effort to increase our precision and push the detection limit for linear spectropolarimetric signatures, we have added a new detector and new LCVR retarders to HiVIS. The detector has bi-directional clocking synchronized with the LCVRs to allow for a two order of magnitude decrease in phase switching time compared to large night-time astronomical spectropolarimeters. Precision in excess of 10$^{-4}$ has been demonstrated in the lab and first light with the instrument has been achieved. With this new mode, we plan to extend our HiVIS survey to include a high-precision component.


\end{document}